\documentclass[aps,
prd,
twocolumn,
superscriptaddress,
nofootinbib
]{revtex4-1}

\usepackage{graphicx}
\usepackage{amssymb}
\usepackage{amsmath}
\usepackage{braket}
\usepackage{listings}
\usepackage{cases}
\usepackage{comment}
\usepackage{cancel}
\usepackage[colorlinks=true,
urlcolor=blue,
anchorcolor=blue,
citecolor=blue,
filecolor=blue,
linkcolor=blue,
menucolor=blue,
linktocpage=true,
pdfproducer=medialab,
pdfa=true
]{hyperref}

\def\slashchar#1{\setbox0=\hbox{$#1$} 
\dimen0=\wd0 
\setbox1=\hbox{/} \dimen1=\wd1 
\ifdim\dimen0>\dimen1 
\rlap{\hbox to \dimen0{\hfil/\hfil}} 
#1 
\else 
\rlap{\hbox to \dimen1{\hfil$#1$\hfil}} 
/ 
\fi}

\newcommand{\ltsim}{\protect\raisebox{-0.5ex}{$\:\stackrel{\textstyle <}{\sim}\:$}}

\newcommand{\dif}[2]{\frac{\mathrm{d} #1}{\mathrm{d} #2}}

\newcommand{\dd}{\mathrm{d}}
\newcommand{\ee}{\mathrm{e}}

\newcommand{\SUSY}{\mathrm{SUSY}}

\begin{document}

\title{PBH Dark Matter in Supergravity Inflation Models}
\date{\today}
\author{Masahiro Kawasaki}
\email{kawasaki@icrr.u-tokyo.ac.jp}
\affiliation{Institute for Cosmic Ray Research, The University of Tokyo, Kashiwa, Chiba 277-8582, Japan}
\affiliation{Kavli Institute for the Physics and Mathematics of the Universe (WPI), UTIAS, The University of Tokyo, Kashiwa, Chiba 277-8583, Japan}
\author{Alexander Kusenko}
\email{kusenko@ucla.edu}
\affiliation{Department of Physics and Astronomy, University of California, Los Angeles, California 90095-1547, USA}
\affiliation{Kavli Institute for the Physics and Mathematics of the Universe (WPI), UTIAS, The University of Tokyo, Kashiwa, Chiba 277-8583, Japan}
\author{Yuichiro Tada}
\email{yuichiro.tada@ipmu.jp}
\affiliation{Institute for Cosmic Ray Research, The University of Tokyo, Kashiwa, Chiba 277-8582, Japan}
\affiliation{Kavli Institute for the Physics and Mathematics of the Universe (WPI), UTIAS, The University of Tokyo, Kashiwa, Chiba 277-8583, Japan}
\author{Tsutomu T. Yanagida}
\email{tsutomu.tyanagida@ipmu.jp}
\affiliation{Kavli Institute for the Physics and Mathematics of the Universe (WPI), UTIAS, The University of Tokyo, Kashiwa, Chiba 277-8583, Japan}

\preprint{IPMU 16-0087}

\begin{abstract}
We propose a novel scenario to produce abundant primordial black holes (PBHs) in new inflation 
which is a second phase of a double inflation in the supergravity frame work.
In our model, some preinflation phase before the new inflation is assumed and it would be responsible for the primordial curvature perturbations
on the cosmic microwave background scale, while the new inflation produces only the small scale perturbations. 
Our new inflation model has linear, quadratic, and cubic terms in its potential and PBH production corresponds with its flat inflection point.
The linear term can be interpreted to come from a supersymmetry-breaking sector, and with this assumption, 
the vanishing cosmological constant condition after inflation and the flatness condition for the inflection point can be consistently satisfied.
\end{abstract}

\maketitle

\section{Introduction}
The primordial black holes (PBHs), which might be formed in the early radiation dominated universe, have attracted scientists for more than 40 years and are recently refocused on more and more.
Theoretically they can be formed by the gravitational collapse of the Hubble patch if the mean energy density in that patch is 
higher by $\mathcal{O}(0.1)$ than its surroundings~\cite{Hawking:1971ei,Carr:1974nx,Carr:1975qj}.
One of the main motivations of PBHs is dark matter (DM). Since they behave as a matter component, they can be a main component of DM without introducing other elementary particles.
However PBHs are still not detected by any observation, and their abundance is 
constrained~\cite{Josan:2009qn,Carr:2009jm,Griest:2013esa,Graham:2015apa,Tisserand:2006zx,Ricotti:2007au,Barnacka:2012bm,Carr:2016drx}, 
except in a mass window of $10^{21\text{--}24}\,\mathrm{g}$, where PBHs can still account for all dark matter. 
Some recent attempts to close this window relied on assumptions that are difficult to justify.  
For example, the existence of neutron stars in globular clusters could help constrain the remaining window, 
if the density of dark matter exceeded the average by more than two orders of magnitude~\cite{Capela:2013yf}.  
However, observations of globular clusters show no evidence of significant dark matter content 
in such systems~\cite{Bradford:2011aq,Ibata:2012eq}.  
Furthermore, it was suggested that tidal deformation of a neutron star could lead to an efficient energy dissipation and capture of a black hole, 
leading to stronger constraints~\cite{Pani:2014rca}, 
but such energy losses are uncertain, and they are likely to be suppressed for realistic parameters and 
velocities in excess of the speed of sound~\cite{Capela:2014qea,Defillon:2014wla}, so that no new constraints can be derived, 
and the window for PBH dark matter remains open. 

As one of the series of the PBH production works, several authors have studied the PBH formation in the double inflation 
in the supergravity frame work~\cite{Kawasaki:1997ju,Kawasaki:1998vx,Kawasaki:2006zv,
Kawaguchi:2007fz,Frampton:2010sw,Kawasaki:2012kn,Kawasaki:2016ijp}.\footnote{For other PBH production models, see Refs. of recent review~\cite{Carr:2016drx}.}
Following this stream, we propose a novel new inflation model as the second phase of the double inflation, 
where PBHs can be produced on mass $\sim10^{22}\,\mathrm{g}$ enough to constitute the bulk of DM in this paper.\footnote{
In the supersymmetric frame work, the lightest supersymmetric particle (LSP) can be a candidate of WIMP DM. Here we assume that
the $R$ parity is broken and LSP is not stable, and then we need PBH DM instead.}
As the preinflation before the new inflation, which is responsible for the curvature perturbations on the CMB scale, any specific model 
is not required to be supposed. The potential of our new inflation model consists of the linear, quadratic, and $n$-th moment terms
and we found that PBHs can be produced on $\sim10^{22}\,\mathrm{g}$ enough to be a main component of DM.
Our model is based on the discrete $R$ symmetry $Z_{2nR}$ and 
we find that only the $n=3$ case could produce the desired spectrum of PBHs. 
It is remarkable that the linear term can be consistently interpreted to come from the supersymmetry (SUSY) breaking sector in the case of $n=3$ 
under the flatness condition at the inflection point and the vanishing cosmological constant condition.
While we propose the abundant PBHs of $\sim10^{22}\,\mathrm{g}$ as DM, 
we can make another peak in the PBH mass spectrum on $\sim30M_\odot$. Those PBHs would have contributed to 
gravitational waves which recently detected by LIGO/Virgo collaboration~\cite{Abbott:2016blz,Abbott:2016nhf} 
as discussed in \cite{Bird:2016dcv,Clesse:2016vqa,Sasaki:2016jop,Eroshenko:2016hmn}.

The rest of this paper is organized as follows. In Section~\ref{new inflation}, we describe the constitution method of new inflation in supergravity  
and introduce our model.
Then we concretely evaluate the current PBH abundance in Section~\ref{PBH mass spectrum}. We make our conclusions in Section~\ref{conclusions}. In Appendix~\ref{multiple horizon crossing},
we describe the treatments for the modes which exit the horizon at near the beginning of the second new inflation.

\section{New inflation with inflection point in supergravity}\label{new inflation}
Let us briefly review the composition of new inflation in supergravity at first.  
Here we adopt the model proposed in Ref.~\cite{Kumekawa:1994gx,Izawa:1997df}. In this model, 
we assume a discrete $R$ symmetry $Z_{2nR}$ which is broken down to a discrete $Z_{2R}$ during and after the inflation. 
The inflaton superfield $\phi$ has an $R$ charge 2.
These assumptions lead the following effective superpotential at the leading order:
\begin{align}
	W(\phi)=v^2\phi-\frac{g}{n+1}\phi^{n+1}.
\end{align}
Here and hereafter we use the Planck units where the reduced Planck mass $M_p\simeq2.4\times10^{18}\,\mathrm{GeV}$ is set to be unity.
The $R$-invariant effective K\"ahler potential can be written as,
\begin{align}
	K(\phi)=|\phi|^2+\frac{\kappa}{4}|\phi|^4+\cdots.
\end{align}
For these super and K\"ahler potentials, the inflaton potential in supergravity is given by,
\begin{align}
	V(\phi)=\ee^K\left(D_\phi WK^{\phi\bar{\phi}}(D_\phi W)^*-3|W|^2\right),
\end{align}
where $D_\phi W=W_\phi+K_\phi W$. Defining the inflaton by the real part of $\phi$, namely $\varphi=\sqrt{2}\,\mathrm{Re}\,\phi$, it can be expanded as,
\begin{align}\label{V of varphi}
	V(\varphi)\simeq v^4-\frac{\kappa}{2}v^4\varphi^2-\frac{g}{2^{\frac{n}{2}-1}}v^2\varphi^n+\frac{g^2}{2^n}\varphi^{2n},
\end{align}
and the slow-roll inflation can be driven either by the quadratic or $n$-th moment term. 
Also it has a negative minimum at $\varphi_\mathrm{min}\simeq\sqrt{2}\left(\frac{v^2}{g}\right)^\frac{1}{n}$ as,
\begin{align}
	\braket{V}\simeq-3\braket{\ee^K|W|^2}\simeq-3\left(\frac{n}{n+1}\right)^2v^4\left(\frac{v^2}{g}\right)^\frac{2}{n}.
\end{align} 
This negative energy would be canceled out after inflation with a positive contribution $\mu_\mathrm{SUSY}^4$ due to a SUSY-breaking effect.
Therefore the gravitino mass can be related with the new inflation scale as,
\begin{align}\label{m32}
	m_{3/2}\simeq\frac{\mu_\mathrm{SUSY}^2}{\sqrt{3}}\simeq\frac{n}{n+1}v^2\left(\frac{v^2}{g}\right)^\frac{1}{n}.
\end{align}
The inflaton mass around the potential minimum is given by,
\begin{align}
	m_\varphi\simeq nv^2\left(\frac{v^2}{g}\right)^{-\frac{1}{n}}.
\end{align}
Therefore, if the inflaton decays into standard model particles simply by Planck suppressed operators, the reheating temperature can be evaluated by,
\begin{align}\label{TR}
	T_R\simeq0.1m_\varphi^{3/2}\simeq0.1n^\frac{3}{2}g^\frac{3}{2n}v^{3-\frac{3}{n}}.
\end{align}
In the next section, we will use this reheating temperature to calculate the PBH mass spectrum.\footnote{
For the parameters which we will use in the next section, the reheating temperature can be estimated as $T_R\sim8.8\times10^8\,\mathrm{GeV}$
with the above assumption. This value may be marginal to realize the thermal leptogenesis~\cite{Fukugita:1986hr}.
However we have checked that desired PBH mass spectra can be achieved even for a higher reheating temperature e.g. 
$T_R\sim10^{10}\mathrm{GeV}$.}

Now let us consider the initial condition for this inflation.
Small field new inflation generally suffers from a severe initial condition problem. 
That is, both the inflaton initial field value and its time derivative should be extremely small to 
have a sufficiently long inflation, but originally there is no reason to stabilize the inflaton field to the potential origin 
since the inflaton potential should be flat enough to satisfy the slow-roll conditions.
Moreover, even if one can introduce some stabilizing term in the potential, new inflation realizes eternal inflation 
if the inflaton's initial field value
is much smaller than the Hubble fluctuation $\frac{H}{2\pi}$ and it should continue much longer than 60 e-folds (we want new inflation to
contribute only to small scale perturbations as we will mention).
As proposed in~\cite{Izawa:1997df}, these problems can be naturally solved in the supergravity frame work 
by introducing a preinflation phase before the new inflation and adding a constant term to superpotential $W_\mathrm{const}=c$.\footnote{Note that, 
in the original model~\cite{Izawa:1997df},
hybrid inflation is assumed as a preinflation, which gives a non-zero superpotential and leads the linear term in the potential of the new inflation. 
However, since we have already introduced the constant term
in the superpotential, the preinflation does not need to give a non-zero superpotential in our model.} 
During the preinflation, the inflaton of the new inflation can have a Hubble induced mass term 
$\frac{1}{2}V_\mathrm{pre}\varphi^2\simeq\frac{3}{2}H^2\varphi^2$ through the coefficient $\ee^K$ of the potential.
Moreover the constant term in the superpotential leads the linear term $2\sqrt{2}cv^2\varphi$ in the potential, which shifts the potential minimum 
from zero to $2\sqrt{2}\frac{cv^2}{V_\mathrm{pre}}$. 
The Hubble induced mass keeps stabilizing the inflaton even after the preinflation until the beginning of the second new inflation 
$V_\mathrm{pre}\simeq v^4$, and therefore the initial field value of $\varphi$ is given by,
\begin{align}
	\varphi_i\simeq2\sqrt{2}\frac{c}{v^2}.
\end{align} 
The new inflation can avoid eternal inflation as long as $\varphi_i$ is sufficiently larger than 
the Hubble fluctuation $\frac{H}{2\pi}$ at the beginning of the new inflation.

Taking above things, we consider the PBH formation in the following model:
\begin{align}
	W_\mathrm{new}=&\, v^2\phi-\frac{g}{n+1}\phi^{n+1}+c, \nonumber \\
	K_\mathrm{new}=&\, |\phi|^2+\frac{\kappa}{4}|\phi|^4+\cdots, \nonumber \\
	V(\varphi)\simeq&\, v^4-2\sqrt{2}cv^2\varphi-\frac{\kappa}{2}v^4\varphi^2-\frac{g}{2^{\frac{n}{2}-1}}v^2\varphi^n
		+\frac{g^2}{2^n}\varphi^{2n} \nonumber \\
	&+V_\mathrm{pre}\left(1+\frac{1}{2}\varphi^2\right).
\end{align}
Here note that, if $\kappa<0$, the linear, quadratic, and $n$-th moment terms can have an inflection point for $\varphi>0$. 
Since $V^\prime$ is locally maximized at that inflection point, 
if the maximum of $V^\prime$ is still negative but quite close to zero, 
the slow-roll inflation are not spoiled and moreover very large curvature perturbations can be produced.
The inflection point $\varphi_*$ can be obtained as,
\begin{align}
	V^{\prime\prime}(\varphi_*)&\simeq|\kappa|v^4-\frac{n(n-1)}{2^{\frac{n}{2}-1}}gv^2\varphi_*^{n-2}=0 \nonumber \\
	\Rightarrow \quad \varphi_*&=\left(\frac{2^{\frac{n}{2}-1}}{n(n-1)}\frac{|\kappa|v^2}{g}\right)^\frac{1}{n-2}.
\end{align}
Then we require the flat inflection condition as,
\begin{align}\label{flat condition}
	V^\prime(\varphi_*)\simeq& -2\sqrt{2}cv^2 \nonumber \\
	&+\frac{n-2}{n-1}\left(\frac{2^{\frac{n}{2}-1}}{n(n-1)}\right)^\frac{1}{n-1}\frac{|\kappa|^\frac{n-1}{n-2}
	v^\frac{4n-6}{n-2}}{g^\frac{1}{n-2}} \sim0, \nonumber \\
	\Rightarrow \quad g\sim&\,\frac{1}{n(n-2)}\left(\frac{n-2}{2(n-1)}\right)^{n-2}\frac{|\kappa|^{n-1}v^{2(n-1)}}{c^{n-2}},
\end{align}
neglecting $V_\mathrm{pre}$ and the higher order term. With this condition, the curvature perturbations generated around $\phi_*$
would be large enough to produce abundant PBHs, and the concrete successful parameters will be shown in the next section.
Here note that the inflection point can be written as $\varphi_*=\frac{n-1}{n-2}\frac{1}{|\kappa|}\varphi_i$ under this condition and therefore
it is automatically set to be slightly larger than the initial field value for $|\kappa|\sim\mathcal{O}(0.1)$. That is because the initial field value $\varphi_i$ is determined by
the balance between the linear term $cv^2\varphi$ and the preinflaiton-induced mass term $v^4\varphi^2$, while the flat inflection is the point where the linear term is comparable with 
the self-induced mass term $\kappa v^4\varphi^2$ which is smaller than the preinflation-induced mass term by a factor $|\kappa|$.

If the new inflation is assumed to realize both small perturbations like as those on the CMB scale and large perturbations which would cause the formation of PBHs, 
it generally takes too many e-folds in the transition from small to large perturbations
since the time derivative of the slow-roll parameter itself is suppressed by the slow-roll parameters, $\dif{}{N}\log\epsilon_V\sim\mathcal{O}(\epsilon_V,\eta_V)$
where $N$ denotes the e-folding number and $\epsilon_V$ and $\eta_V$ represent the slow-roll parameters $\frac{1}{2}\left(\frac{V^\prime}{V}\right)^2$ and $\frac{V^{\prime\prime}}{V}$ respectively.
However, since we have already introduced double inflation, the new inflation can be free from the COBE normalization by simply assuming that the preinflation is responsible for the CMB scale perturbations, 
and then the new inflation can end in sufficiently short time in that case. Indeed we will show in the next section that there are parameter regions where PBHs can be produced enough to be a main component of DM
and the constraints for large scale perturbations like the CMB spectral distortion can be avoided.
Here if one assume that the curvature perturbations generated in the new inflation are large even apart from the inflection point,
the linear term itself is required to be small enough. 
Letting $A_\mathrm{lin}$ denote the amplitude of the power spectrum
of the curvature perturbations during the linear term in the potential dominantly contributes to the perturbations, 
the constant superpotential $c$ is determined through the following relation:
\begin{align}\label{Alin}
	A_\mathrm{lin}=\frac{1}{12\pi^2}\frac{V^3}{V^{\prime2}}\simeq\frac{1}{96\pi^2}\frac{v^8}{c^2}, 
	\quad \Leftrightarrow \quad c\simeq\frac{1}{\sqrt{96\pi^2A_\mathrm{lin}}}v^4.
\end{align}
Therefore, for $A_\mathrm{lin}\sim10^{-3}$, $c$ is required to be as small as $v^4$.

The above two conditions~(\ref{flat condition}) and (\ref{Alin}) are required for PBH formation. Now combining them clarifies the $v$-dependence of $g$ as $g\sim|\kappa|^{n-1}v^{6-2n}$.
Therefore, in the case of $n=3$, $g\sim|\kappa|^2$ does not depend on the new inflation scale $v$ and could be smaller than unity.  
On the other hand, for $n\ge4$, $g$ becomes much larger than unity for $v\ll1$ and such a large $g$ would spoil unitarity of the theory 
for example~\cite{Harigaya:2013pla}.
Moreover for large $n$ the duration of the new inflation after the inflection point will be short due to its steep potential 
and it tends to make the PBH mass small. Indeed we have checked that the PBH mass spectrum are tilted to
the lighter mass and conflicts with the constraints on the PBH abundance for $n\ge4$.
For those reasons, we will concentrate on the case of $n=3$ hereafter.\footnote{In addition, 
$n=3$ is uniquely favored by the anomaly free conditions for supersymmetric standard gauge groups
with the discrete $R$ symmetry $Z_{2nR}$~\cite{Evans:2011mf}.}

As an interesting fact, the small constant term $c$ in the superpotential can be interpreted to come from the SUSY-breaking sector 
in the case of $n=3$~\cite{Takahashi:2013cxa}.\footnote{Note that the case of $n=4$ is considered in Ref.~\cite{Takahashi:2013cxa}
since their motivation is not to produce PBHs but to modify the spectral index in new inflation and therefore the required condition is different.}
The SUSY-breaking F-term order $\mu_\mathrm{SUSY}^2$ naively arise from the term like,
\begin{align}
	W_{\cancel{\mathrm{SUSY}}}=\mu_\mathrm{SUSY}^2Z.
\end{align}
If $Z$ obtains a vev $\braket{Z}\sim\mu_\mathrm{SUSY}$, this term can lead the constant superpotential 
$c\sim\mu_\mathrm{SUSY}^3$. Indeed it can be realized in the dynamical SUSY breaking models proposed 
in Ref.~\cite{Izawa:1996pk,Intriligator:1996pu}
if the origin of $Z$ is destabilized due to a large Yukawa coupling ($\sim4\pi$), and the estimated constant term is given by 
$c\sim\frac{\Lambda^3}{(4\pi)^2}$ where the dynamical scale $\Lambda$ is related with $\mu_\mathrm{SUSY}$ by 
$\mu_\mathrm{SUSY}^2=\frac{\Lambda^2}{4\pi}$. On the other hand, under the flat inflection condition~(\ref{flat condition})
and the large curvature perturbation condition~(\ref{Alin}), the vanishing cosmological constant condition~(\ref{m32}) gives the 
parameter dependence of the SUSY-breaking scale as $\mu_\SUSY\sim|\kappa|^\frac{1-n}{2n}v^{2\frac{n-1}{n}}$, neglecting numerical factors.
Therefore the scale dependence of the constant term $c$ is consistent with the above assumption if and only if $n=3$ as,
\begin{align}\label{condition for c}
	c\sim v^4\sim\mu_\SUSY^3.
\end{align}
It can be checked that this consistency is retained for the concrete parameter values which we will show in the next section
even if numerical factors are included.

\section{Formation of primordial black holes}\label{PBH mass spectrum}
\begin{figure}
	\centering
	\includegraphics[width=0.9\hsize]{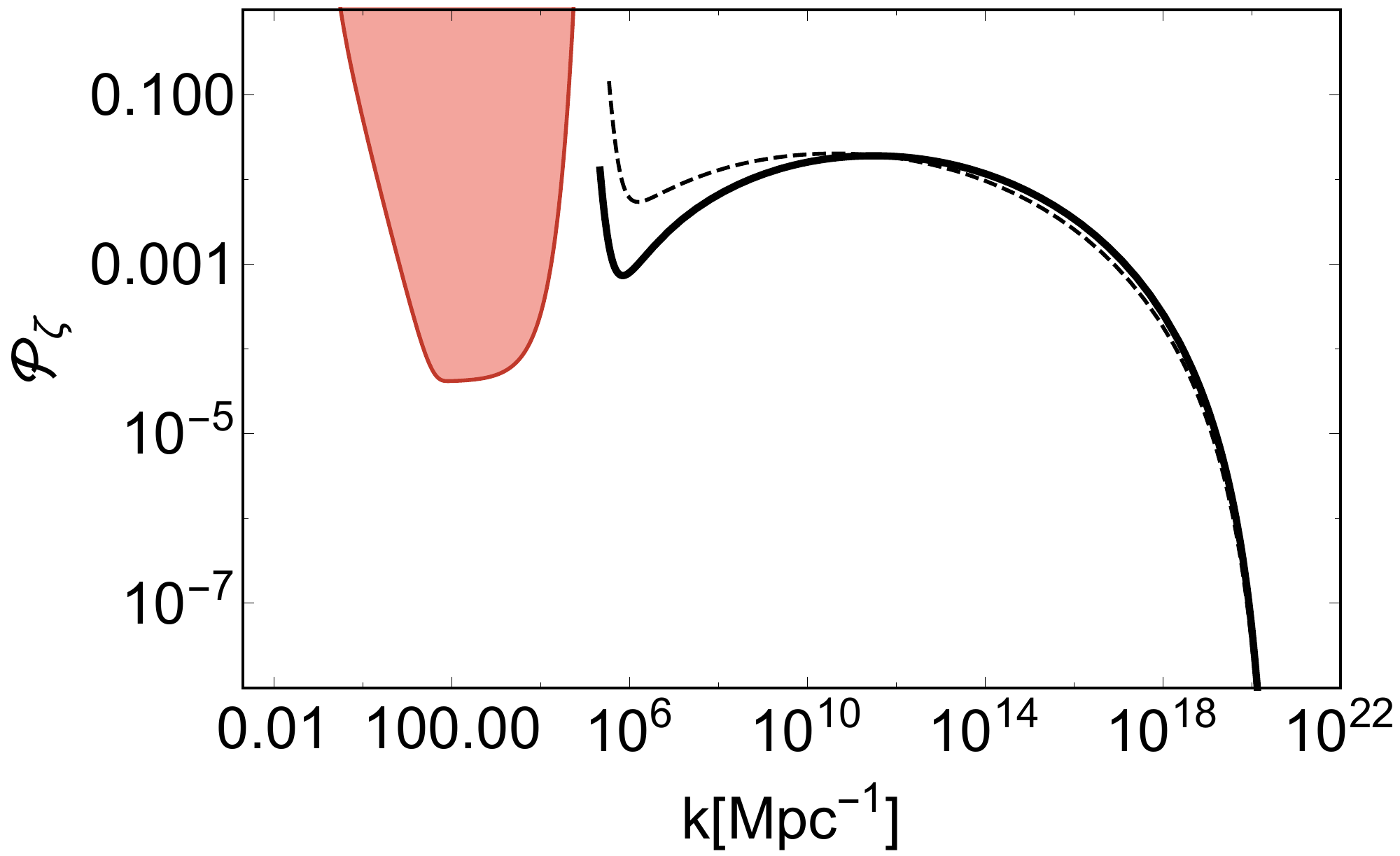}
	\caption{The power spectra of the curvature perturbations for parameters (\ref{single peak parameter}) (solid line)
	and (\ref{DP parameter}) (dashed line). The red line represents the constraints on the power spectrum from the non-detection of the CMB
	$\mu$-distortion estimated by Eq.~(\ref{mu-dist}). The modes $k\ltsim10^6\,\mathrm{Mpc^{-1}}$ actually reenter the horizon between 
	the two inflations
	and the treatments for them are described in Appendix~\ref{multiple horizon crossing}.}
	\label{power}
\end{figure}

In this section, we will calculate the PBH abundance and exemplify parameter sets where PBHs constitute the main component of DM.
At first, given the inflation scale $v$ and the reheating temperature $T_R$, the perturbation scale can be related with the backward e-folds by~\cite{Lyth:2009zz},
\begin{align}
	&\log\left(\frac{k}{0.002\,\mathrm{Mpc}^{-1}}\right) \nonumber \\
	&\simeq-N+56+\frac{2}{3}\log\left(\frac{v}{10^{16}\,\mathrm{GeV}}\right)+\frac{1}{3}\log\left(\frac{T_R}{10^9\,\mathrm{GeV}}\right).
\end{align}
We will use Eq.~(\ref{TR}) as the reheating temperature hereafter. Then, with the potential~(\ref{V of varphi}) and the Hubble induced mass $\frac{1}{2}V_\mathrm{pre}\varphi^2$, 
we can calculate the power spectrum of the curvature perturbations
$\mathcal{P}_\zeta(k)$ in the standard linear theory and the result is shown in Fig.~\ref{power} as the black thick line 
for the following parameter values:
\begin{align}\label{single peak parameter}
	v\!=\!10^{-4}, \, \kappa\!=\!-0.44, \, c\!=\!1.838\!\times\!10^{-16}, \, g\!=\!5.036\!\times\!10^{-3}.
\end{align}
In this calculation, we have simply assumed that the preinflation potential behaves 
as a matter component $V_\mathrm{pre}\propto a^{-3}$ after the preinflation.
For these parameters, the vanishing cosmological constant~(\ref{m32}) and the SUSY-breaking assumption~(\ref{condition for c})
consistently predict $m_{3/2}\sim10^8\,\mathrm{GeV}$ which suggests the pure gravity mediation~\cite{Ibe:2011aa}.
In this figure, we also show the constraints from the CMB spectral $\mu$-distortion as the red region. 
The CMB $\mu$-distortion from the Silk damping of a single $k$-mode can be approximated by~\cite{Kohri:2014lza},
\begin{align}\label{mu-dist}
	\mu\sim2.2\mathcal{P}_\zeta(k)\left[\exp\left[-\frac{k_{\mathrm{Mpc}^{-1}}}{5400}\right]
	-\exp\left[-\left(\frac{k_{\mathrm{Mpc}^{-1}}}{31.6}\right)^2\right]\right],
\end{align}
where $k_{\mathrm{Mpc}^{-1}}$ represent the wavenumber in $\mathrm{Mpc}^{-1}$.
We have used the current 2$\sigma$ upper limit $\mu\ltsim9\times10^{-5}$ by the COBE/FIRAS experiment~\cite{Fixsen:1996nj}.
Finally, the modes $k\ltsim10^6\,\mathrm{Mpc^{-1}}$ actually reenter the horizon between the preinflation and 
the second new inflation. We will describe the treatments of these modes in Appendix~\ref{multiple horizon crossing}.

With use of this power spectrum, the PBH abundance can be calculated as follows. 
At first, the mass of PBH is almost given by the horizon mass when the overdensity reenters the horizon, 
and let $\gamma$ denote the ratio between them here. That is, the PBH mass corresponding with the scale $k$ is given by,
\begin{align}
	M_\mathrm{PBH}(k)&=\left.\gamma\frac{4\pi}{3}\rho H^{-3}\right|_{k=aH}
	=\gamma M_\mathrm{eq}\left(\frac{g_{*\mathrm{eq}}}{g_*}\right)^{1/2}\left(\frac{T_\mathrm{eq}}{T}\right)^2 \nonumber \\
	&=\gamma M_\mathrm{eq}\left(\frac{g_{*\mathrm{eq}}}{g_*}\right)^{1/6}\left(\frac{k_\mathrm{eq}}{k}\right)^2,
\end{align}
where $M_\mathrm{eq}$ denotes the horizon mass at the matter-radiation equality calculated as,
\begin{align}
	M_\mathrm{eq}=\frac{4\pi}{3}\rho_\mathrm{eq}H_\mathrm{eq}^{-3}=\frac{8\pi}{3}\frac{\rho_r^0}{a_\mathrm{eq}k_\mathrm{eq}^3}.
\end{align}
Also we have used an approximation that the effective d.o.f. for energy density $g_*$ is almost equal to that for entropy density $g_{*s}$.
Using $\rho_r^0=7.84\times10^{-34}\,\mathrm{g\,cm^{-3}}$, $k_\mathrm{eq}=0.07\Omega_mh^2\,\mathrm{Mpc^{-1}}$, $a_\mathrm{eq}^{-1}=24000\Omega_mh^2$, and $g_{*\mathrm{eq}}=3.36$,
we can finally obtain~\cite{Josan:2009qn},
\begin{align}\label{MPBH}
	\hspace{-10pt}M_\mathrm{PBH}(k)\!=\!3.6M_\odot\left(\frac{\gamma}{0.2}\right)\!\left(\!\frac{g_*|_{k=aH}}{106.75}\!\right)^{-1/6}\!
	\!\left(\!\frac{k}{10^6\,\mathrm{Mpc^{-1}}}\!\right)^{-2}\!\hspace{-10pt},
\end{align}
where $M_\odot\simeq2\times10^{33}\,\mathrm{g}$ is the solar mass. In the simple analytic calculation~\cite{Carr:1975qj}, $\gamma$ is evaluated as $\gamma=3^{-3/2}\simeq0.2$ and 
we will use this value hereafter.

\begin{figure}
	\centering
	\includegraphics[width=0.9\hsize]{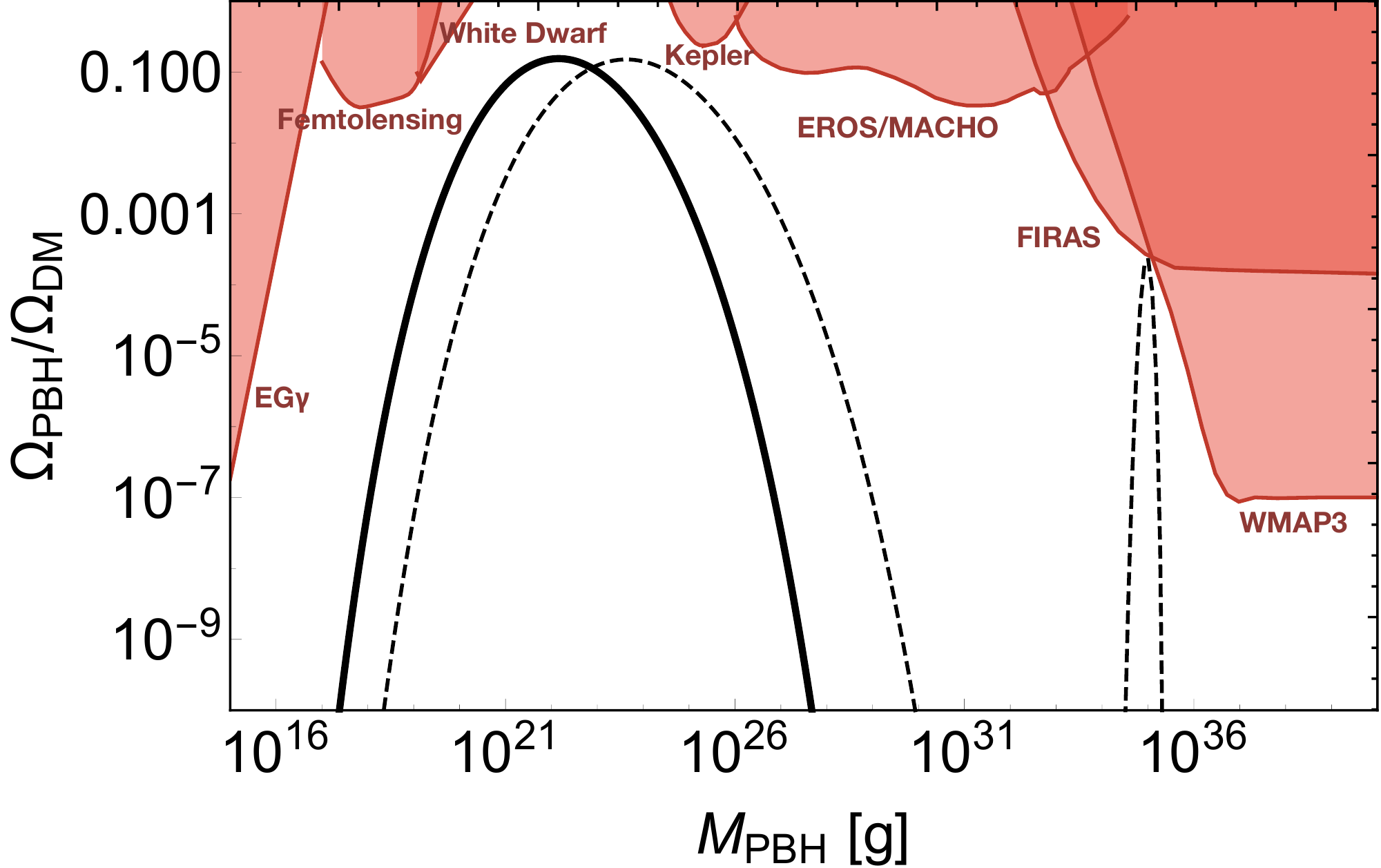}
	\caption{The PBH fraction to DM in each logarithmic mass bin with the parameters~(\ref{single peak parameter}) (solid line) and
	(\ref{DP parameter}) (dashed line). Several constraints for PBH abundance are also shown as red 
	regions: from left to right, extragalactic $\gamma$-rays from evapolation~\cite{Carr:2009jm,Carr:2016drx}, femtolensing of $\gamma$-ray bursts~\cite{Barnacka:2012bm},
	existence of white dwarfs in our local galaxy~\cite{Graham:2015apa}, Kepler microlensing and millilensing~\cite{Griest:2013esa}, EROS/MACHO microlensing~\cite{Tisserand:2006zx},
	and accretion effects on CMB~\cite{Ricotti:2007au}.
	For both the solid and dashed lines, 
	the total integrated fraction is about unity, therefore these models predicts abundant PBHs as a main component of DM. Moreover in the
	dashed line case there is another peak at $\sim30M_\odot$, which might cause GW150914~\cite{Sasaki:2016jop,Eroshenko:2016hmn}.}
	\label{fPBH}
\end{figure}

The formation rate of PBHs $\beta$ is given by the probability of excess over the threshold. 
That is, under the assumption that the density perturbations follow the Gaussian distribution,\footnote{
Note that the non-Gaussianity (NG) is expected to be small since the second new inflation is driven almost only by a single scalar field
around the inflection point. Indeed we have briefly checked that the local non-linearity parameter $f_\mathrm{NL}$ is as small as $\sim0.1$
and for such a small NG it is known that the predicted PBH abundance is hardly affected~\cite{Saito:2008em,Byrnes:2012yx,Young:2013oia}.
However the second peak on $^\sim30M_\odot$ which we will show later might be modified by NG effects since it corresponds with
the phase of the beginning of the new inflation. We leave this problem for future works.}
\begin{align}
	\beta(M_\mathrm{PBH}) \!=\! \int_{\delta_c}\dd\delta\frac{1}{\sqrt{2\pi\sigma^2(k)}}\ee^{-\frac{\delta^2}{2\sigma^2(k)}}
	\!=\!\frac{1}{2}\mathrm{Erfc}\!\left(\!\frac{\delta_c}{\sqrt{2}\sigma(k)}\!\right)\!.
\end{align}
Here $\delta_c$ represents the threshold density perturbation and we will adopt the simple analytic estimation $\delta_c=p_r/\rho_r=1/3$~\cite{Carr:1975qj}. $\sigma^2(k)$ is the variance of the comoving density perturbations
coarse-grained on $k^{-1}$, which is given by~\cite{Young:2014ana},
\begin{align}
	\sigma^2(k)=\int\dd\log k^\prime\, W^2(k^\prime/k)\frac{16}{81}(k^\prime/k)^4\mathcal{P}_\zeta(k).
\end{align}
$W(z)$ represents the Fourier transformed window function and we will adopt the Gaussian window $W(z)=\ee^{-z^2/2}$ in the following calculations.
Note that the PBH mass and coarse-graining scale $k$ are related by Eq.~(\ref{MPBH}). 
The formation rate directly gives the ratio of the PBH energy density to the total energy density 
at the horizon reentering, $\beta(M_\mathrm{PBH})=\left.\rho_\mathrm{PBH}/\rho\right|_{k=aH}$. Therefore the current PBH fraction to
DM for a single mass mode can be derived as,
\begin{align}
	f_\mathrm{PBH}(M_\mathrm{PBH})=&\,\frac{\Omega_\mathrm{PBH}h^2}{\Omega_\mathrm{DM}h^2}
	=\frac{\Omega_mh^2}{\Omega_\mathrm{DM}h^2}\left.\frac{\rho_\mathrm{PBH}(M_\mathrm{PBH})}{\rho}\right|_\mathrm{eq} \nonumber \\
	=&\,\frac{\Omega_mh^2}{\Omega_\mathrm{DM}h^2}\frac{T}{T_\mathrm{eq}}\beta(M_\mathrm{PBH}) \nonumber \\
	=&\,1.3\times10^8\beta(M_\mathrm{PBH})\left(\frac{\Omega_\mathrm{DM}h^2}{0.12}\right)^{-1}\left(\frac{\gamma}{0.2}\right)^{1/2} \nonumber \\
	&\times\left(\frac{g_*}{106.75}\right)^{-1/4}\left(\frac{M_\mathrm{PBH}}{M_\odot}\right)^{-1/2},
\end{align}
where $\Omega_\mathrm{DM}h^2$ represents the current DM abundance $\Omega_\mathrm{DM}h^2\sim0.12$~\cite{Ade:2015xua}. 
The resultant PBH fraction is plotted in Fig.~\ref{fPBH} as the black thick line.
We also show several observational constraints for PBH abundance as red regions~\cite{Carr:2009jm,Griest:2013esa,Graham:2015apa,Tisserand:2006zx,Ricotti:2007au,Barnacka:2012bm,Carr:2016drx}.
Here we have not used the constraints from existence of neutron stars~\cite{Capela:2013yf,Pani:2014rca} and white dwarfs~\cite{Capela:2012jz} in globular clusters since 
they require the high DM-density assumption whose validity seems questionable
as mentioned in introduction.
Although the shown PBH fraction does not reach unity in each logarithmic mass bin,
the total PBH fraction:
\begin{align}
	f_\mathrm{PBH,tot}=\int\dd\log M\,f_\mathrm{PBH}(M),
\end{align}
reaches unity with the current parameters. Therefore our model can describe the formation of sufficient PBHs to be a main component of DM.
One might think that the PBH mass spectrum can be shifted to the more massive direction and even the neutron stars constraints could be avoided for some parameters.
However we have checked that the $\mu$-distortion constraints shown in Fig.~\ref{power} as the red line cannot be satisfied in that case.

Before ending this section, let us concentrate on the sharp peaks of the power spectrum around $k\sim10^6\,\mathrm{Mpc^{-1}}$ shown in Fig.~\ref{power}. 
This peak comes from the fact that at the beginning of the new inflation the Hubble induced mass 
$\frac{1}{2}V_\mathrm{pre}\varphi^2$
cannot be neglected yet and the inflaton potential is being slightly flattened due to its stabilizing effect at first. For a slightly smaller $c$ (namely a slightly flatter linear potential), this peak can be enlarged and
another peak of the fraction for more massive PBHs can be shown. For the following parameters:
\begin{align}\label{DP parameter}
	v\!=\!10^{-4}, \, \kappa\!=\!-0.182, \, c\!=\!5.53\!\times\!10^{-17}, \, g\!=\!4.26\!\times\!10^{-3},
\end{align}
the resultant power spectrum and PBH fraction are plotted in Fig.~\ref{power} and \ref{fPBH} as the black dashed lines. 
Again the total PBH fraction is about unity ($f_\mathrm{PBH,tot}\simeq1$) for them,
but they show another peak at $M_\mathrm{PBH}\sim30M_\odot$. Recently LIGO/Virgo collaboration succeeded in the first direct detection of gravitational waves GW150914, 
which came from an inspiral and merger of a black hole binary~\cite{Abbott:2016blz}. The masses of these black holes are estimated as $36^{+5}_{-4}M_\odot$ and $29^{+4}_{-4}M_\odot$.
Taking this, the binary formation rate of PBHs whose masses are around $30M_\odot$ has been evaluated by several 
authors~\cite{Bird:2016dcv,Clesse:2016vqa,Sasaki:2016jop,Eroshenko:2016hmn},
and the authors of \cite{Sasaki:2016jop} claimed that the binary formation rate of PBHs 
satisfying current constraints ($f_\mathrm{PBH}(\sim30M_\odot)\ltsim10^{-4}$) would be high enough to be consistent with 
the black hole merger rate $2\text{--}400\,\mathrm{Gpc^{-3}yr^{-1}}$ inferred from LIGO observations~\cite{Abbott:2016nhf} 
(however it was claimed in \cite{Eroshenko:2016hmn} that the binary formation rate might be smaller than that evaluated in \cite{Sasaki:2016jop}).
Therefore our model might explain both of DM whose main component is as $\sim10^{24}\,\mathrm{g}$ PBHs and 
GW150914 from the merger of the $\sim30M_\odot$ PBH binary.\footnote{Recently, in Ref.~\cite{Cheng:2016qzb}, it has also been proposed that
$30M_\odot$-PBH could be produced by the gauge field production with the Chern-Simons coupling $\phi F_{\mu\nu}\tilde{F}^{\mu\nu}$.} 

However, the existence of the peak around $30M_{\odot}$ in our model should be examined
more carefully since the corresponding fluctuation modes reenter the horizon 
between the preinflation and the new inflation and in such a case effects of the metric perturbations 
that are not included in the present analysis might be important . 
So we need to solve full evolutions of fluctuations of scalar fields and metric perturbations, which 
will be investigated in future work.

\section{Conclusions}\label{conclusions}
In this paper, we proposed a new inflation model as a second phase of a double inflation 
consistently constituted in supergravity frame work, 
where sufficient primordial black holes (PBHs) can be produced to be a main component of dark matter (DM).
Any specific inflationary model is not required for the preinflation which is responsible for the large scale curvature perturbations,
as long as it is consistent with the observations of, e.g., CMB.
The potential of the new inflation in our model consists of the linear, quadratic, and cubic terms and has a flat inflection point where PBHs can be produced.
The specific power spectra of the curvature perturbations and the resultant PBH mass spectra are shown in Fig.~\ref{power} and \ref{fPBH} for two parameter sets~(\ref{single peak parameter}) and (\ref{DP parameter}).
The inflection point corresponds with $\sim10^{22}\,\mathrm{g}$ PBHs and they constitute the bulk of DM in our model. In addition, we can make another peak for the PBH mass spectrum on $\sim30M_\odot$
as indicated by the black dashed line in Fig.~\ref{fPBH}. Such a peak corresponds with the beginning of the new inflation and those PBHs might cause the gravitational waves detected by LIGO/Virgo collaboration.

\begin{acknowledgements}
We thank Keisuke Inomata for useful comments and discussion.
This work is supported by MEXT KAKENHI Grant Number 15H05889 (M.K.), 16H02176 (T.T.Y.), 
JSPS KAKENHI Grant Number 25400248 (M.K.), 26104009 and 26287039 (T.T.Y.),  
and also by the World Premier International Research Center Initiative (WPI), MEXT, Japan.
A.K. is supported by the U. S. Department of Energy Grant DE-SC0009937.
Y.T. is supported by JSPS Research Fellowship for Young Scientists. 
\end{acknowledgements}

\appendix

\section{Multiple horizon crossing modes}\label{multiple horizon crossing}
In this appendix, we will describe the treatments for the modes which exit the horizon at the end of the preinflation once, enter the horizon 
between two inflations, and then reexit the horizon at the beginning of the second new inflation. They correspond with the modes for $k\ltsim10^6\,\mathrm{Mpc^{-1}}$ in our parameters.
Their dynamics are non-trivial due to their multiple horizon crossing and the amplitudes of them at the second horizon exit which determine
the curvature perturbations in the new inflation have to be evaluated carefully.
To do so, one must solve EoM for perturbations continuously over the two inflations and connection phase, 
including the effects of the metric perturbations.
However here let us roughly estimate them in the super- and subhorizon limit without metric perturbations.

The linear EoM for perturbations which have a Hubble induced mass $\frac{3}{2}H^2\varphi^2$ is given by,
\begin{align}
	0&=\delta\ddot{\varphi}+3H\delta\dot{\varphi}+\left(3H^2+\frac{k^2}{a^2}\right)\delta\varphi \nonumber \\
	&\sim
	\begin{cases} 
		\displaystyle
		\delta\ddot{\varphi}+3H\delta\dot{\varphi}+3H^2\delta\varphi, & k\ll aH, \\
		\displaystyle
		\delta\ddot{\varphi}+3H\delta\dot{\varphi}+\frac{k^2}{a^2}\delta\varphi, & k\gg aH.
	\end{cases}
\end{align}
In the second line, we have used the super- and subhorizon limit. The subhorizon EoM can be always rewritten, with use of the
conformal time $a\dd\eta=\dd t$ and in the subhorizon limit, as,
\begin{align}
	\partial_\eta^2(a\delta\varphi)+\frac{k^2}{a^2}(a\delta\varphi)\simeq0,
\end{align}
and therefore it only has oscillating solutions whose amplitudes decrease as $a^{-1}$. On the other hand, assuming that the background EoS
is given by $\omega=p/\rho>-1$ ($a\propto t^\frac{2}{3(1+\omega)}$), the superhorizon EoM reads,
\begin{align}
	\delta\ddot{\varphi}+\frac{2}{(1+\omega)t}\delta\dot{\varphi}+\frac{4}{3(1+\omega)^2t^2}\delta\varphi=0.
\end{align}
It can be easily solved by assuming the power-law solution $t^n$ and the real part of the power is given by 
$\mathrm{Re}[n]=\frac{-1+3\omega}{2(1+\omega)}$. That is, the amplitude of the solutions damps as 
$t^\frac{-1+\omega}{2(1+\omega)}\propto a^{-\frac{3(1-\omega)}{4}}$. Also, in the exact de Sitter background, the Hubble parameter is constant
and the two solutions are soon found as $\delta\varphi\propto\exp\left(-\frac{3}{2}Ht\pm\frac{\sqrt{3}}{2}iHt\right)$. Namely the perturbations damp
as $a^{-3/2}$ (therefore the damping factor $a^{-\frac{3(1-\omega)}{4}}$ for $\omega>-1$ can be used also for the de Sitter case in which $\omega=-1$).
In summary, the amplitude of the perturbations decreases as $a^{-1}$ in the subhorizon limit 
and as $a^{-\frac{3(1-\omega)}{4}}$ in the superhorizon limit.	

\begin{figure}
	\centering
	\includegraphics[width=0.9\hsize]{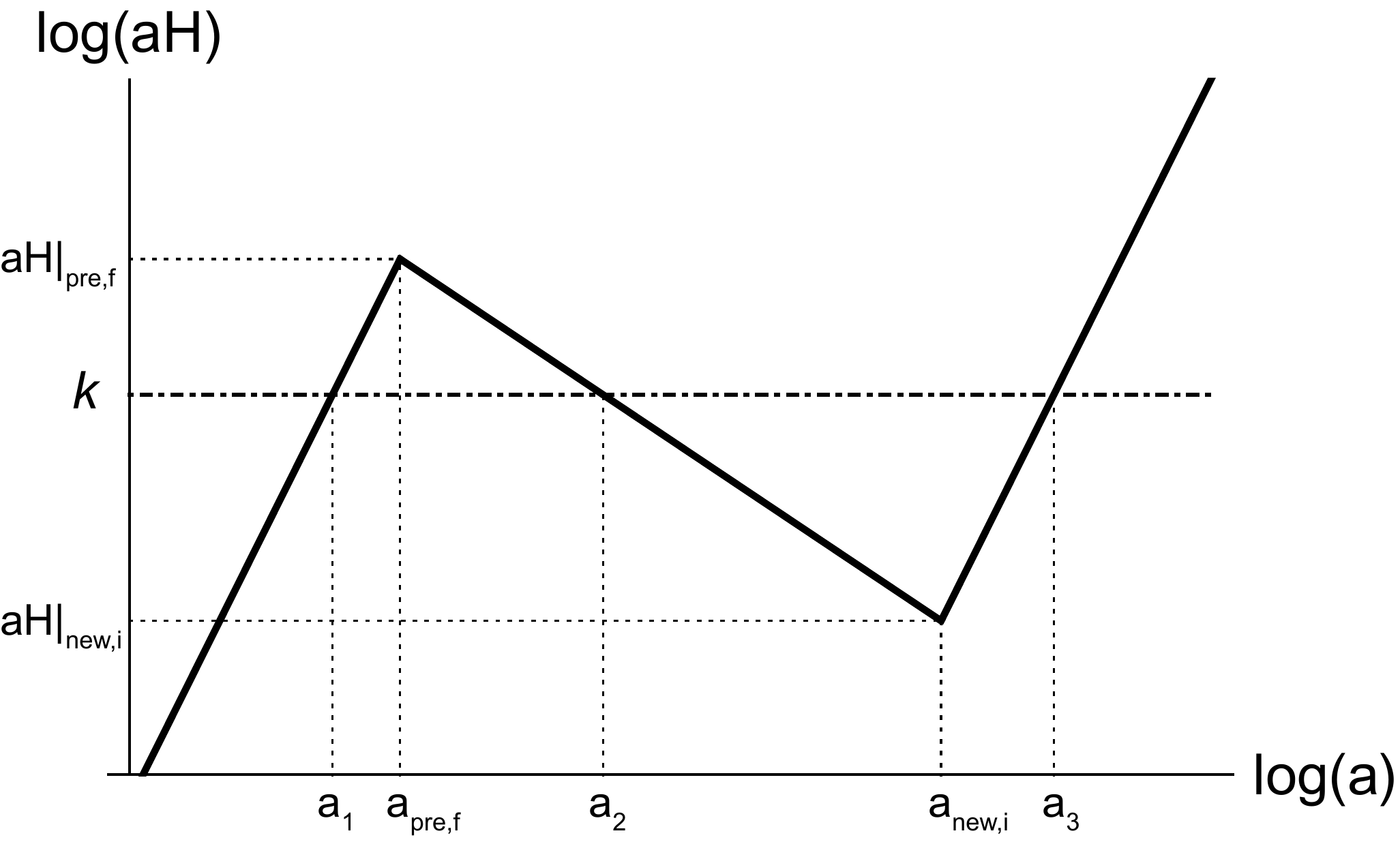}
	\caption{The schematic image of the relation between the horizon scale $aH$ and the multiple horizon crossing mode $k$. The time before $a_\mathrm{pre,f}$ corresponds with that during the preinflation, while
	$a_\mathrm{new,i}$ represents the beginning of the new inflation. Between the two inflations, the horizon scale generally decreases as $aH\propto a^{-\frac{1+3\omega}{2}}$ with $\omega=p/\rho$.
	Therefore the modes which exit the horizon at near the end of the preinflation can reenter the horizon during the preinflaton oscillation phase, and then reexit the horizon at the beginning of the new inflation.
	Those modes contribute to the sharp peaks of the power spectra around $k\sim10^6\,\mathrm{Mpc^{-1}}$ shown in Fig.~\ref{power}, which can lead the second peak to the PBH fraction on $\sim30M_\odot$.
	In the super or subhorizon limits, it can be shown that the amplitude of the perturbations at the second horizon exit $a_3$ is equal to the standard value $\frac{H_\mathrm{new}}{2\pi}$ in spite of their complicated
	horizon crossing process.}
	\label{aH}
\end{figure}

Now let us evaluate the concrete amplitude of the multiple horizon crossing modes at the horizon exit during the new inflation. 
Letting $aH|_\mathrm{pre,f}$ and $aH|_\mathrm{new,i}$ denote the horizon scale at the end of the preinflation and the beginning of the new inflation, such modes correspond with $aH|_\mathrm{new,i}<k<aH|_\mathrm{pre,f}$.
We illustrate an schematic image about the relation between the wavelength $k$ and the horizon scale $aH$ in Fig.~\ref{aH} as the solid and 
dot-dashed lines respectively.
At the first horizon exit during the preinflation, the perturbation amplitude is given by the standard one $\delta\varphi\sim\frac{H_\mathrm{pre}}{2\pi}$. 
After that, the amplitude decreases until the second horizon exit during the new inflation as,
\begin{align}
	\left.\delta\varphi\right|_3\sim&\,\frac{H_\mathrm{pre}}{2\pi}\left(\frac{a_\mathrm{pre,f}}{a_1}\right)^{-3/2}
	\left(\frac{a_2}{a_\mathrm{pre,f}}\right)^{-\frac{3(1-\omega)}{4}} \nonumber \\
	&\times\left(\frac{a_\mathrm{new,i}}{a_2}\right)^{-1}\left(\frac{a_3}{a_\mathrm{new,i}}\right)^{-1},
\end{align}
following the scale factor dependences which we showed previously. 
Here the subscript 1, 2, and 3 represent each horizon crossing time during the preinflation, between two inflations, and during the new inflation respectively. Also we have referred to the EoS
between two inflations as $\omega$. Noting that the horizon scale $aH$ is proportional to $a^{-\frac{1+3\omega}{2}}$, it can be rewritten as,
\begin{align}
	\left.\delta\varphi\right|_3\sim&\,\frac{H_\mathrm{pre}}{2\pi}\left(\frac{aH|_\mathrm{pre,f}}{k}\right)^{-3/2}
	\left(\frac{k}{aH|_\mathrm{pre,f}}\right)^\frac{3(1-\omega)}{2(1+3\omega)} \nonumber \\
	&\times\left(\frac{aH|_\mathrm{new,i}}{k}\right)^\frac{2}{1+3\omega}\left(\frac{k}{aH|_\mathrm{new,i}}\right)^{-1} \nonumber \\
	=&\,\frac{H_\mathrm{pre}}{2\pi}\left(\frac{aH|_\mathrm{pre,f}}{aH|_\mathrm{new,i}}\right)^{-\frac{3(1+\omega)}{1+3\omega}}.
\end{align}
Finally, using $aH\propto a^{-\frac{1+3\omega}{2}}\propto\rho^\frac{1+3\omega}{6(1+\omega)}$, we can obtain,
\begin{align}
	\left.\delta\varphi\right|_3\sim\frac{H_\mathrm{pre}}{2\pi}\left(\frac{\rho_\mathrm{pre}}{\rho_\mathrm{new}}\right)^{-1/2}\simeq\frac{H_\mathrm{new}}{2\pi}.
\end{align}
That is, the resultant amplitude is eventually identical with the standard one even though they experienced a complicated process.
	
The results obtained in this appendix would be changed if the effects of, e.g., the metric perturbations or the resonance.
For example, if one assumes the mass term chaotic inflation as the preinflation, the Hubble induced mass 
$\frac{1}{2}V_\mathrm{pre}\varphi^2=\frac{1}{4}m^2\chi^2\varphi^2$ where $\chi$ represents the inflaton of the preinflation, leads
the parametric resonance~\cite{Kofman:1994rk}. However this problem would be solved by introducing a K\"ahler coupling $K\supset|X|^2|\phi|^2$ where
$X$ is the preinflaton superfield, since this K\"ahler potential brings $\dot{\chi}^2\varphi^2$ coupling for example and it cancels and reduces 
the oscillation of the Hubble induced mass. Also the metric perturbations would also grow density perturbations 
on subhorizon scales~\cite{Jedamzik:2010dq,Easther:2010mr}. 
However the peak of the PBH mass spectrum on
$\sim30M_\odot$ is mainly caused by the largest scale mode $k\sim aH|_\mathrm{new,i}$ which is hardly affected by the subhorizon effect.
Moreover, even if the amplitude of the perturbations would be slightly modified, it could be absorbed into the parameter tuning.
Anyway more strict analysis for those modes is postponed to future works.



\begin{thebibliography}{99}


\bibitem{Hawking:1971ei} 
  S.~Hawking,
  Mon.\ Not.\ Roy.\ Astron.\ Soc.\  {\bf 152}, 75 (1971).
  
\bibitem{Carr:1974nx} 
  B.~J.~Carr and S.~W.~Hawking,
  Mon.\ Not.\ Roy.\ Astron.\ Soc.\  {\bf 168}, 399 (1974).
  
\bibitem{Carr:1975qj} 
  B.~J.~Carr,
  Astrophys.\ J.\  {\bf 201}, 1 (1975).
  doi:10.1086/153853
  
  

\bibitem{Josan:2009qn} 
  A.~S.~Josan, A.~M.~Green and K.~A.~Malik,
  Phys.\ Rev.\ D {\bf 79}, 103520 (2009)
  doi:10.1103/PhysRevD.79.103520
  [arXiv:0903.3184 [astro-ph.CO]].
  
\bibitem{Carr:2009jm} 
  B.~J.~Carr, K.~Kohri, Y.~Sendouda and J.~Yokoyama,
  Phys.\ Rev.\ D {\bf 81}, 104019 (2010)
  doi:10.1103/PhysRevD.81.104019
  [arXiv:0912.5297 [astro-ph.CO]].

\bibitem{Tisserand:2006zx} 
  P.~Tisserand {\it et al.} [EROS-2 Collaboration],
  Astron.\ Astrophys.\  {\bf 469}, 387 (2007)
  doi:10.1051/0004-6361:20066017
  [astro-ph/0607207].
  
\bibitem{Ricotti:2007au} 
  M.~Ricotti, J.~P.~Ostriker and K.~J.~Mack,
  Astrophys.\ J.\  {\bf 680}, 829 (2008)
  doi:10.1086/587831
  [arXiv:0709.0524 [astro-ph]].

\bibitem{Barnacka:2012bm} 
  A.~Barnacka, J.~F.~Glicenstein and R.~Moderski,
  Phys.\ Rev.\ D {\bf 86}, 043001 (2012)
  doi:10.1103/PhysRevD.86.043001
  [arXiv:1204.2056 [astro-ph.CO]].
  
\bibitem{Griest:2013esa} 
  K.~Griest, A.~M.~Cieplak and M.~J.~Lehner,
  Phys.\ Rev.\ Lett.\  {\bf 111}, no. 18, 181302 (2013).
  doi:10.1103/PhysRevLett.111.181302
  
\bibitem{Graham:2015apa} 
  P.~W.~Graham, S.~Rajendran and J.~Varela,
  Phys.\ Rev.\ D {\bf 92}, no. 6, 063007 (2015)
  doi:10.1103/PhysRevD.92.063007
  [arXiv:1505.04444 [hep-ph]].
  


\bibitem{Carr:2016drx} 
  B.~Carr, F.~Kuhnel and M.~Sandstad,
  arXiv:1607.06077 [astro-ph.CO].


  
\bibitem{Capela:2013yf} 
  F.~Capela, M.~Pshirkov and P.~Tinyakov,
  Phys.\ Rev.\ D {\bf 87}, no. 12, 123524 (2013)
  doi:10.1103/PhysRevD.87.123524
  [arXiv:1301.4984 [astro-ph.CO]].
  
       


\bibitem{Bradford:2011aq} 
  J.~D.~Bradford {\it et al.},
  Astrophys.\ J.\  {\bf 743}, 167 (2011)
  Erratum: [Astrophys.\ J.\  {\bf 778}, 85 (2013)]
  doi:10.1088/0004-637X/743/2/167, 10.1088/0004-637X/778/1/85
  [arXiv:1110.0484 [astro-ph.CO]].
  
\bibitem{Ibata:2012eq} 
  R.~Ibata, C.~Nipoti, A.~Sollima, M.~Bellazzini, S.~Chapman and E.~Dalessandro,
  Mon.\ Not.\ Roy.\ Astron.\ Soc.\  {\bf 428}, 3648 (2013)
  doi:10.1093/mnras/sts302
  [arXiv:1210.7787 [astro-ph.CO]].
  
\bibitem{Pani:2014rca} 
  P.~Pani and A.~Loeb,
  JCAP {\bf 1406}, 026 (2014)
  doi:10.1088/1475-7516/2014/06/026
  [arXiv:1401.3025 [astro-ph.CO]].
  
\bibitem{Capela:2014qea} 
  F.~Capela, M.~Pshirkov and P.~Tinyakov,
  arXiv:1402.4671 [astro-ph.CO].
  
\bibitem{Defillon:2014wla} 
  G.~Defillon, E.~Granet, P.~Tinyakov and M.~H.~G.~Tytgat,
  Phys.\ Rev.\ D {\bf 90}, no. 10, 103522 (2014)
  doi:10.1103/PhysRevD.90.103522
  [arXiv:1409.0469 [gr-qc]].
  
  

\bibitem{Kawasaki:1997ju} 
  M.~Kawasaki, N.~Sugiyama and T.~Yanagida,
  Phys.\ Rev.\ D {\bf 57}, 6050 (1998)
  doi:10.1103/PhysRevD.57.6050
  [hep-ph/9710259].

\bibitem{Kawasaki:1998vx} 
  M.~Kawasaki and T.~Yanagida,
  Phys.\ Rev.\ D {\bf 59}, 043512 (1999)
  doi:10.1103/PhysRevD.59.043512
  [hep-ph/9807544].
  
\bibitem{Kawasaki:2006zv} 
  M.~Kawasaki, T.~Takayama, M.~Yamaguchi and J.~Yokoyama,
  Phys.\ Rev.\ D {\bf 74}, 043525 (2006)
  doi:10.1103/PhysRevD.74.043525
  [hep-ph/0605271].
  
\bibitem{Kawaguchi:2007fz} 
  T.~Kawaguchi, M.~Kawasaki, T.~Takayama, M.~Yamaguchi and J.~Yokoyama,
  Mon.\ Not.\ Roy.\ Astron.\ Soc.\  {\bf 388}, 1426 (2008)
  doi:10.1111/j.1365-2966.2008.13523.x
  [arXiv:0711.3886 [astro-ph]].
  
\bibitem{Frampton:2010sw} 
  P.~H.~Frampton, M.~Kawasaki, F.~Takahashi and T.~T.~Yanagida,
  JCAP {\bf 1004}, 023 (2010)
  doi:10.1088/1475-7516/2010/04/023
  [arXiv:1001.2308 [hep-ph]].
  
\bibitem{Kawasaki:2012kn} 
  M.~Kawasaki, A.~Kusenko and T.~T.~Yanagida,
  Phys.\ Lett.\ B {\bf 711}, 1 (2012)
  doi:10.1016/j.physletb.2012.03.056
  [arXiv:1202.3848 [astro-ph.CO]].
 
\bibitem{Kawasaki:2016ijp} 
  M.~Kawasaki, K.~Mukaida and T.~T.~Yanagida,
  arXiv:1605.04974 [hep-ph].
    
 
    
    

   
  
  

\bibitem{Abbott:2016blz} 
  B.~P.~Abbott {\it et al.} [LIGO Scientific and Virgo Collaborations],
  Phys.\ Rev.\ Lett.\  {\bf 116}, no. 6, 061102 (2016)
  doi:10.1103/PhysRevLett.116.061102
  [arXiv:1602.03837 [gr-qc]].
  
\bibitem{Abbott:2016nhf} 
  B.~P.~Abbott {\it et al.} [LIGO Scientific and Virgo Collaborations],
  arXiv:1602.03842 [astro-ph.HE].
  
  
  

\bibitem{Bird:2016dcv} 
  S.~Bird, I.~Cholis, J.~B.~Munoz, Y.~Ali-Haimoud, M.~Kamionkowski, E.~D.~Kovetz, A.~Raccanelli and A.~G.~Riess,
  Phys.\ Rev.\ Lett.\  {\bf 116}, no. 20, 201301 (2016)
  doi:10.1103/PhysRevLett.116.201301
  [arXiv:1603.00464 [astro-ph.CO]].
  
\bibitem{Clesse:2016vqa} 
  S.~Clesse and J.~Garc\'ia-Bellido,
  arXiv:1603.05234 [astro-ph.CO].
  
\bibitem{Sasaki:2016jop} 
  M.~Sasaki, T.~Suyama, T.~Tanaka and S.~Yokoyama,
  arXiv:1603.08338 [astro-ph.CO].
  
\bibitem{Eroshenko:2016hmn} 
  Y.~N.~Eroshenko,
  arXiv:1604.04932 [astro-ph.CO].
  
  
  
  

\bibitem{Kumekawa:1994gx} 
  K.~Kumekawa, T.~Moroi and T.~Yanagida,
  Prog.\ Theor.\ Phys.\  {\bf 92}, 437 (1994)
  doi:10.1143/PTP.92.437
  [hep-ph/9405337].

\bibitem{Izawa:1997df} 
  K.~I.~Izawa, M.~Kawasaki and T.~Yanagida,
  Phys.\ Lett.\ B {\bf 411}, 249 (1997)
  doi:10.1016/S0370-2693(97)01040-X
  [hep-ph/9707201].
  


\bibitem{Fukugita:1986hr} 
  M.~Fukugita and T.~Yanagida,
  Phys.\ Lett.\ B {\bf 174}, 45 (1986).
  doi:10.1016/0370-2693(86)91126-3
  
  

\bibitem{Harigaya:2013pla} 
  K.~Harigaya, M.~Ibe and T.~T.~Yanagida,
  Phys.\ Rev.\ D {\bf 89}, no. 5, 055014 (2014)
  doi:10.1103/PhysRevD.89.055014
  [arXiv:1311.1898 [hep-ph]].
  
  


\bibitem{Evans:2011mf} 
  J.~L.~Evans, M.~Ibe, J.~Kehayias and T.~T.~Yanagida,
  Phys.\ Rev.\ Lett.\  {\bf 109}, 181801 (2012)
  doi:10.1103/PhysRevLett.109.181801
  [arXiv:1111.2481 [hep-ph]].
  


\bibitem{Takahashi:2013cxa} 
  F.~Takahashi,
  Phys.\ Lett.\ B {\bf 727}, 21 (2013)
  doi:10.1016/j.physletb.2013.10.026
  [arXiv:1308.4212 [hep-ph]].

\bibitem{Izawa:1996pk} 
  K.~I.~Izawa and T.~Yanagida,
  Prog.\ Theor.\ Phys.\  {\bf 95}, 829 (1996)
  doi:10.1143/PTP.95.829
  [hep-th/9602180].
  
\bibitem{Intriligator:1996pu} 
  K.~A.~Intriligator and S.~D.~Thomas,
  Nucl.\ Phys.\ B {\bf 473}, 121 (1996)
  doi:10.1016/0550-3213(96)00261-1
  [hep-th/9603158].
  
  
  

\bibitem{Lyth:2009zz} 
  D.~H.~Lyth and A.~R.~Liddle,
  Cambridge, UK: Cambridge Univ. Pr. (2009) 497 p
  
  
  

\bibitem{Ibe:2011aa} 
  M.~Ibe and T.~T.~Yanagida,
  Phys.\ Lett.\ B {\bf 709}, 374 (2012)
  doi:10.1016/j.physletb.2012.02.034
  [arXiv:1112.2462 [hep-ph]].
  
  

\bibitem{Kohri:2014lza} 
  K.~Kohri, T.~Nakama and T.~Suyama,
  Phys.\ Rev.\ D {\bf 90}, no. 8, 083514 (2014)
  doi:10.1103/PhysRevD.90.083514
  [arXiv:1405.5999 [astro-ph.CO]].
    
    



\bibitem{Fixsen:1996nj} 
  D.~J.~Fixsen, E.~S.~Cheng, J.~M.~Gales, J.~C.~Mather, R.~A.~Shafer and E.~L.~Wright,
  Astrophys.\ J.\  {\bf 473}, 576 (1996)
  doi:10.1086/178173
  [astro-ph/9605054].
    
    

\bibitem{Saito:2008em} 
  R.~Saito, J.~Yokoyama and R.~Nagata,
  JCAP {\bf 0806}, 024 (2008)
  doi:10.1088/1475-7516/2008/06/024
  [arXiv:0804.3470 [astro-ph]].

\bibitem{Byrnes:2012yx} 
  C.~T.~Byrnes, E.~J.~Copeland and A.~M.~Green,
  Phys.\ Rev.\ D {\bf 86}, 043512 (2012)
  doi:10.1103/PhysRevD.86.043512
  [arXiv:1206.4188 [astro-ph.CO]].
  
\bibitem{Young:2013oia} 
  S.~Young and C.~T.~Byrnes,
  JCAP {\bf 1308}, 052 (2013)
  doi:10.1088/1475-7516/2013/08/052
  [arXiv:1307.4995 [astro-ph.CO]].
    
   
\bibitem{Young:2014ana} 
  S.~Young, C.~T.~Byrnes and M.~Sasaki,
  JCAP {\bf 1407}, 045 (2014)
  doi:10.1088/1475-7516/2014/07/045
  [arXiv:1405.7023 [gr-qc]].

  
  





 


 
\bibitem{Ade:2015xua} 
  P.~A.~R.~Ade {\it et al.} [Planck Collaboration],
  arXiv:1502.01589 [astro-ph.CO].
 
 
 

\bibitem{Capela:2012jz} 
  F.~Capela, M.~Pshirkov and P.~Tinyakov,
  Phys.\ Rev.\ D {\bf 87}, no. 2, 023507 (2013)
  doi:10.1103/PhysRevD.87.023507
  [arXiv:1209.6021 [astro-ph.CO]].
 


\bibitem{Cheng:2016qzb} 
  S.~L.~Cheng, W.~Lee and K.~W.~Ng,
  arXiv:1606.00206 [astro-ph.CO].
  
  
  

  
 

\bibitem{Kofman:1994rk} 
  L.~Kofman, A.~D.~Linde and A.~A.~Starobinsky,
  Phys.\ Rev.\ Lett.\  {\bf 73}, 3195 (1994)
  doi:10.1103/PhysRevLett.73.3195
  [hep-th/9405187].
  
\bibitem{Jedamzik:2010dq} 
  K.~Jedamzik, M.~Lemoine and J.~Martin,
  JCAP {\bf 1009}, 034 (2010)
  doi:10.1088/1475-7516/2010/09/034
  [arXiv:1002.3039 [astro-ph.CO]].
  
\bibitem{Easther:2010mr} 
  R.~Easther, R.~Flauger and J.~B.~Gilmore,
  JCAP {\bf 1104}, 027 (2011)
  doi:10.1088/1475-7516/2011/04/027
  [arXiv:1003.3011 [astro-ph.CO]].
    
  
 

  
 
  
  

  
  

  
  

  
  
  
  

\end{thebibliography}
\end{document}